\documentclass{article}
\usepackage{amsmath}
\usepackage{authblk}
\usepackage{caption}
\usepackage{cite}
\usepackage{color}
\usepackage{soul}
\usepackage{float}
\usepackage{geometry}
\usepackage{graphicx}
\usepackage{hyperref}
\usepackage{listings}
\usepackage{textcomp}
\usepackage{epstopdf}
\usepackage{hyperref}

\numberwithin{equation}{section}

\lstdefinelanguage{partialsagemathmaxima}
{
	morekeywords=
	{
		reset, from, import, var, assume, print, show, inverse, display, expr, function, for, range, len, diff, numerator, exp, factor, full_simplify, rhs, collect_common_factors, solve, def, abs, while,break,return,hypergeometric,plot, simplify_full,kill,if,get,then,load,ratwtlvl,ratfac,matrix,invert,else,in,expand,collect,save,subs,
	},
	sensitive=false,
	morecomment=[l]{\#},
}
\title{Symbolic analysis of second-order ordinary differential equations with polynomial coefficients}
\author{Tolga Birkandan\thanks{E-mail: birkandant@itu.edu.tr} \\
	Department of Physics, Istanbul Technical University, 34469 Istanbul,
	Turkey.}
\begin{document}
	\maketitle
\begin{abstract}
The singularity structure of a second-order ordinary differential equation with polynomial coefficients often yields the type of solution. It is shown that the $\theta$-operator method can be used as a symbolic computational approach to obtain the indicial equation and the recurrence relation. Consequently, the singularity structure leads to the transformations that yield a solution in terms of a special function, if the equation is suitable. Hypergeometric and Heun-type equations are mostly employed in physical applications. Thus only these equations and their confluent types are considered with SageMath routines which are assembled in the open-source package symODE2.
\end{abstract}

\section{Introduction} 
Mathematical analysis of physical problems generally requires the methods of solving ordinary differential equations (ODEs). Numerical solutions of the initial and boundary value problems are often sufficient to give a concrete idea about the behavior of the system, whereas the analytic solution of an ODE in closed form, especially in terms of special functions may have more importance than constituting an exact solution. In some systems, the type of solution has the potential to reveal the symmetries of the system. For example, the emergence of the hypergeometric function in the solutions may indicate the conformal symmetry \cite{Hortacsu:2001bp,mirjam1,mirjam2,mirjam3}.

The applications of the general and confluent hypergeometric equation have dominated the 20th-century \cite{dere}. Although the mathematical theory of the Heun equation and its confluent forms are far from complete, these functions have been known and employed by experts in the area for many years \cite{Ronveaux,Slavyanov}. Besides, the number of their applications increased substantially after the implementation of the Heun functions in the computer algebra system Maple \cite{maple} in 2005 \cite{Hortacsu,Birkandan:2007ey,Birkandan:2006ac,fiziev06}. The implementation of the Heun type functions in Mathematica in 2020 is another big leap for the Heun community \cite{mathematica,mathplot}.

Most of the free and open-source computer algebra systems and packages involve the solutions of the hypergeometric equation, its confluent form, and related equations at least numerically \cite{sage,maxima,reduce}. However, the symbolic solutions of the Heun type equations are defined only in Maple and Mathematica which are commercial systems. The numerical evaluation of the general Heun and singly confluent Heun functions are studied by Motygin using the freely available GNU Octave language \cite{octave,oleg1,oleg2}. The recent work by Giscard and Tamar also deals with the numerical calculation of the Heun type functions \cite{giscard}. The work is in progress for numerical treatment of the Heun type equations under Python, and the initial results of the work are presented in \cite{numericalheun}.

A historical review on computer algebra systems is given in \cite{MacCallum:2018csx} along with gravitational applications. The general methods for obtaining symbolic solutions to ODEs and a review of comprehensive literature before the year 2000 are given in \cite{grabmeier}. Among the papers released prior to 2000, we should cite the seminal paper of Kovacic \cite{kovacic} and Duval and Loday-Richaud's work in which the hypergeometric and Heun type equations are studied, in particular \cite{duval}. Among the papers on the solutions of ODEs in terms of special functions that are published after 2000, we can cite Bronstein and Lafaille \cite{brons}, Chan and Cheb-Terrab \cite{chan}, and van Hoeij with his collaborators \cite{hoe1,hoe2,hoe3}. The extended form of the Nikiforov-Uvarov method is also a powerful tool for analyzing Heun-type equations \cite{karayer1,karayer2,karayer3}.

SageMath is a free and open-source general-purpose computer algebra system licensed under the GPL \cite{sage}. SageMath offers a Python-based language and it is built on many open-source packages such as Maxima, SciPy, NumPy, and matplotlib. A variety of modules are present for many areas such as differential geometry and tensor calculus \cite{sageman1,sageman2} that enable calculations on quantum field theory and general relativity \cite{bebeli}. 

We will focus on the singularity analysis and symbolic solutions of the hypergeometric and Heun type equations using the SageMath system and present the open-source package symODE2 which allows users to analyze these equations symbolically without using a commercial program.

This paper is organized in the following way: In the second section, we present the code structure of our package and give a comparison with the existing codes. In the third and fourth sections, we briefly explain the analysis of the singularity structure and series solutions for a second order ODE with polynomial coefficients, respectively. In the fifth section, we explain our approach for finding the symbolic solutions of the hypergeometric and Heun type equations. Section six involves our conclusions and we describe the standard forms of the equations in the appendix.

\section{The code structure of the symODE2 package}
The symODE2 package is written under SageMath 9.1 using a laptop computer with Intel(R) Core(TM) i7-6500U CPU @ 2.50GHz and 8 GB memory. The operating system is Windows 10 Enterprise ver.1909. It is also tested under SageMath 9.2.

The package consists of two main parts:
\begin{itemize}
	\item \verb|ode2analyzer.sage| for the general analysis and,
	\item \verb|hypergeometric_heun.sage| for the symbolic solutions of the equations. 
\end{itemize}
\verb|hypergeometric_heun.sage| calls the routines defined in \verb|ode2analyzer.sage| when needed. 

We suggest the user put these two files in the same directory. The parts of the package and a sample worksheet can be downloaded from the address \cite{mygithub}:

\url{https://github.com/tbirkandan/symODE2}

The User Manual which can be found at the same address contains detailed explanations of the routines and the analysis of the cases in the sample SageMath worksheet.

\subsection{General analysis (ode2analyzer):}
The first part, \verb|ode2analyzer| contains the routines that
\begin{itemize}
	\item[--] finds the singularity structure of the input ODE. The output is an array that involves the locations of the singularities, indices of the regular singularities, and the ranks of the irregular singularities. 
	\item[--] finds the indices and/or the recurrence relation with respect to a regular singular point using the $\theta$-operator method which will be defined below.
	\item[--] performs a change of variables.
	\item[--] finds the normal form of a second-order ODE. For an input in the form (\ref{firsteq}), the output is in the form (\ref{normalform}).
\end{itemize}

\subsection{Symbolic solutions of special ODEs (hypergeometric\_heun):}
The second part, \verb|hypergeometric_heun| contains the routines that
\begin{itemize}
	\item[--] finds the type of the ODE using its singularity structure and solves it using the routines defined below.
	\item[--] uses a change of variables list in order to bring the input ODE into a special form that is recognized in the package.
	\item[--] solves a hypergeometric equation.	
	\item[--] solves a confluent hypergeometric equation.
	\item[--] solves a general Heun equation.
	\item[--] solves a (singly) confluent Heun equation.
	\item[--] solves a double confluent Heun equation.
	\item[--] solves a biconfluent Heun equation.
	\item[--] solves a triconfluent Heun equation.
\end{itemize}
\subsection{Comparison with the existing codes}
symODE2 is the first freely available, open-source package that allows a symbolic treatment of the Heun-type equations and it is written on SageMath which is also a freely available, open-source program. Therefore, symODE2 provides a free alternative to the commercial programs Maple and Mathematica when the problem is expressing the solutions of these equations symbolically.

symODE2 uses the internal functions of SageMath for the numerical analysis of the hypergeometric-type equations. The numerical treatment of the Heun-type equations is not implemented in SageMath. However, this work is in progress under Python in order to reach a larger community, and SageMath will be able to use the Python code directly \cite{numericalheun}. The commercial programs Maple and Mathematica provide numerical operations as well as symbolics for both hypergeometric and Heun-type equations. Maple and Mathematica also provide the derivatives of the Heun-type functions, unlike symODE2.

The Heun-type equations are implemented in Mathematica in 2020. Figure (\ref{fig:fig2}) that will be given in Section (\ref{sersolsec}) below is created using symODE2 and it can be used as a comparison with the Mathematica results as explained.

The Maple implementation of the Heun-type equations goes back to 2005 and many of the Heun-related applications in the literature published after this year are likely to be done by employing this program. The literature-based cases given in the sample SageMath worksheet \cite{mygithub} and described in the User Manual show that the results obtained by symODE2 agree with the ones in the literature.

\section{Singularity analysis}
A second order ODE can be written in the form,
\begin{equation}\label{firsteq}
	f_1(x)\frac{d^2y(x)}{dx^2}+f_2(x)\frac{dy(x)}{dx}+f_3(x)y(x)=0.
\end{equation}
We can denote $p(x)=f_2(x)/f_1(x)$ and $q(x)=f_3(x)/f_1(x)$ to obtain,
\begin{equation}
	\frac{d^2y}{dx^2}+p(x)\frac{dy}{dx}+q(x)y=0. \label{cici}
\end{equation}
If the functions $p(x)$ and $q(x)$ are analytic at a point $x=x_0$, then $x_0$ is an ``ordinary point" for this ODE. 

The points that make $p(x)$ or $q(x)$ non-analytic are called the singular points or singularities of the ODE. If $x_*$ is a singular point and if $(x-x_*)p(x)$ and $(x-x_*)^2 q(x)$ are both analytic at $x=x_*$, then $x_*$ is called a ``regular singular point". Otherwise, the singular point is ``irregular" \cite{olver}. The singularity behavior at $x \rightarrow \infty$ can be analyzed by performing the transformation $\tilde x=1/x$ and checking the behavior at $\tilde x=0$. If all the singular points of an ODE are regular, then the ODE is said to be a ``Fuchsian equation".

If the singularity at $x=x_*$ is irregular but  $(x-x_*)^k p(x)$ and $(x-x_*)^{2k} q(x)$ are analytic where $k$ is the least integer satisfying this condition, then the irregular singular point at $x=x_*$ has a rank $(k-1)$ \cite{olver}. Consequently, a regular singular point is of rank-$0$ as $k=1$.

We can write the equation (\ref{cici}) in normal form in which the coefficient of the first derivative vanishes. We define $y(x)=g(x)w(x)$ and for
\begin{equation}
	g(x)=e^{-\frac{1}{2}\int_{}^{x} p(x')dx'},
\end{equation}
we obtain
\begin{equation}\label{normalform}
	\frac{d^2w}{dx^2}+\bar{q}(x)w=0,
\end{equation}
where
\begin{equation}
	\bar{q}(x)=q(x)-\frac{1}{2} \frac{dp(x)}{dx}-\frac{p(x)^2}{4}.
\end{equation}
Our code transforms the input equation into the normal form in order to deal with the singular points of only one function, namely $\bar{q}(x)$. Although the package involves a routine for general change of variables, the results of the transformation $\tilde x=1/x$ is included in the function that finds the singularity behavior as the analysis of $x \rightarrow \infty$ case is inevitable.

\section{Series solution around a regular singular point}\label{sersolsec}
An indicial equation can be defined for a regular singular point \cite{Slavyanov}. For a finite regular singular point $x_*$ we have,
\begin{equation}
	r(r-1)+p_* r+q_*=0,
\end{equation}
where $p_*$ and $q_*$ are the residues of $p(x)$ and $(x-x_*)q(x)$ at $x=x_*$, respectively. For the regular singularity at infinity, one can write
\begin{equation}
	r(r+1)-p_\infty r+q_\infty=0,
\end{equation}
where $p_\infty$ and $q_\infty$ are the residues of $p(x)$ and $xq(x)$ at $x \rightarrow \infty$, respectively.

The roots $r=r_1$ and $r=r_2$ of the indicial equation are called the ``indices" or ``characteristic exponents", or ``Frobenius exponents" of the corresponding regular singularity \cite{Slavyanov}. The sum of the all ($2n$) indices corresponding to all ($n$) regular singular points in a Fuchsian equation should be equal to $n-2$ \cite{maierheuntohyper}.

One can find at least one series solution, the ``Frobenius solution" of the form,
\begin{equation}\label{sersol}
	y=\sum_{n=0}^{\infty}C_n (x-x_*)^{n+r},
\end{equation}
near a finite regular singular point $x_*$, $r$ being a characteristic exponent associated with $x_*$. The details on the second solution and the solution around infinity can be found in  \cite{Slavyanov}. We substitute the solution (\ref{sersol}) into equation (\ref{firsteq}) to obtain a recurrence relation among the coefficients $C_n$. For example, the hypergeometric equation admits a two-term recurrence relation which connects $C_{n}$ with $C_{n-1}$, while the Heun equation has a three-term recurrence relation connecting $C_{n}$, $C_{n-1}$ and $C_{n-2}$.

The $\theta$-operator method yields the indicial equation and the recurrence relation for a regular singular point with less effort than the formal Frobenius series calculation. Following \cite{nasa}, we define $D=\frac{d}{dx}$ and,
\begin{eqnarray}
	\theta=x\frac{d}{dx}=xD,\\
	\theta(\theta-1)=x^2D^2.	
\end{eqnarray}
Similarly, we have
\begin{equation}\label{derivs}
	x^nD^n=(-1)^n(-\theta)_n,
\end{equation}
where
\begin{equation}
	(-\theta)_n=\prod_{j=0}^{n-1}(-\theta+j),
\end{equation}
is the generalized factorial notation. A general $n^{th}$ order ODE can be written as
\begin{equation}
	[a_0(x)D^n+a_1(x)D^{n-1}+...+a_n(x)]y=0.
\end{equation}
Using eq.(\ref{derivs}), we can write the equation in the $\theta$-form, namely
\begin{equation}
	[A_0(\theta)+xA_1(\theta)+x^2A_2(\theta)+...+x^mA_m(\theta)]y=0,
\end{equation}
if the coefficients of the original equation are polynomials in $x$. Here, $A_{0,1,...,m}(\theta)$ are polynomials in $\theta$. 

Let us assume that $x=0$ is a regular singular point and seek a series solution in the form (\ref{sersol}) for $x_*=0$. It is  known that any polynomial expression in $\theta$ operating on $x^n$ yields the same polynomial in $n$ times $x^n$, thus $P(\theta)x^n=P(n)x^n$ \cite{nasa}. Using this property, we get
\begin{equation}
	\sum_{n=0}^{\infty}C_n[A_0(n+r) x^{n+r}+A_1(n+r) x^{n+r+1}+...+A_m(n+r) x^{n+r+m}]=0.
\end{equation}
In order to have an arbitrary $C_0$, the indicial equation is obtained as $A_0(r)=0$. Let us shift the indices in the sum,
\begin{equation}
	\sum_{n=0}^{\infty}[C_n A_0(n+r)+C_{n-1} A_1(n+r-1)+...+C_{n-m} A_m(n+r-m)]x^{n+r}=0,
\end{equation}
and the recurrence relation reads
\begin{equation}
	C_n A_0(n+r)+C_{n-1} A_1(n+r-1)+...+C_{n-m} A_m(n+r-m)=0.
\end{equation}
For a regular singular point $x_*$ other than zero, one should make the transformation $x'=x-x_*$ and do the calculation for $x'$.
For the details of the $\theta$-operator method, proofs and examples, we refer the reader to \cite{nasa}.

In our code, the function that finds the indices and the recurrence relation for a given regular singular point is based on the $\theta$-operator method.

\begin{figure}[H]
	\centering
	\includegraphics[scale=0.6]{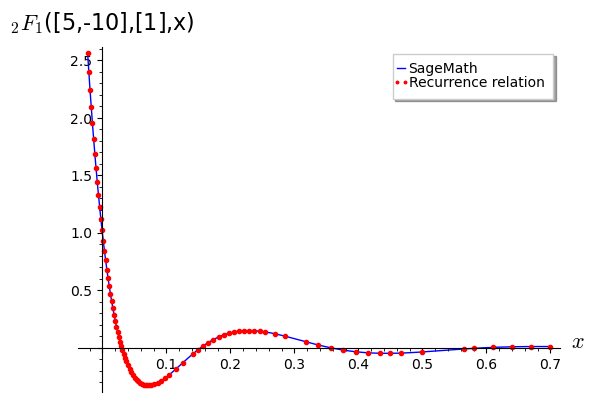} 
	\caption{Plot of the hypergeometric function found by our recurrence relation result and the internal command of SageMath.}
	\label{fig:fig1}%
\end{figure}

The hypergeometric function is implemented in SageMath as \verb|hypergeometric([a,b],[c],x)|. Therefore we can verify our recurrence relation graphically as seen in Figure (\ref{fig:fig1}).
We should note that finding a solution using the recurrence relation generally requires more effort than presented here. Methods such as analytic continuation should be carefully applied in order to deal with the circle of convergence of the series solution \cite{oleg1,oleg2}.

\begin{figure}[H]
	\centering
	\includegraphics[scale=0.6]{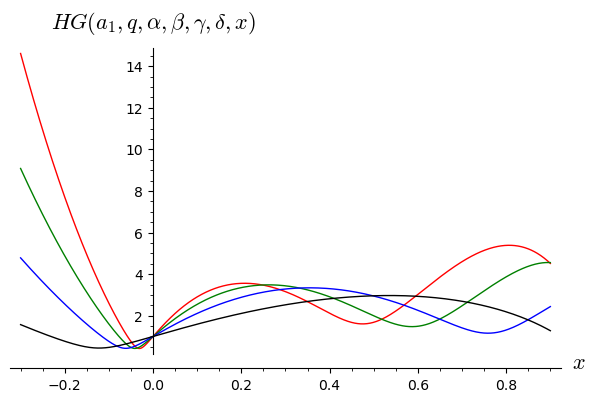} 
	\caption{Plot of the general Heun function similar to the one given in \cite{mathplot}.}
	\label{fig:fig2}%
\end{figure}

We can also plot the series solution for the general Heun equation in Figure (\ref{fig:fig2}) using a similar code with an array of plots and compare it with the plot given in the Wolfram Blog post \cite{mathplot} to see that they are similar. The details of this analysis can be found in the User Manual and the sample SageMath session \cite{mygithub}.

\section{Hypergeometric and Heun-type equations}
The code attempts to find symbolic solutions of some special ODES, namely, the hypergeometric equation, the Heun equation, and their confluent forms. The analysis of the equations is based on the singularity structure. Locations of the singularities and corresponding characteristic exponents play a major role in the method. Using particular substitutions and transformations, the input equation is brought into a standard form that can be recognized by the routines.

A basic example of our approach can be given by using the hypergeometric equation. The Riemann $P$-symbol for the standard form of the hypergeometric equation (\ref{dlmfhyper}) is
\begin{equation}
	P\begin{Bmatrix}0&1&\infty&\\
		0&0&a&x\\
		1-c&c-a-b&b&\end{Bmatrix}.
\end{equation}
Here, the locations of the singular points are given in the first row and each column exhibits the characteristic exponents of the corresponding singular points as found by our code in the sample worksheet. 

Let us study a more general second order ODE with three regular singular points ($x_1,x_2,x_3$) and corresponding indices ($c_{i1}, c_{i2}$, $i=1,2,3$), namely,
\begin{equation}\label{key}
	P\begin{Bmatrix}x_1&x_2&x_3&\\
		c_{11}&c_{21}&c_{31}&x\\
		c_{12}&c_{22}&c_{32}&\end{Bmatrix}.
\end{equation}
The substitution,
\begin{equation}
	u(x) \rightarrow
	\begin{cases}
		\big(\frac{x-x_1}{x-x_3} \big)^{c_{11}}\big(\frac{x-x_2}{x-x_3} \big)^{c_{21}}u(x), & \text{if } x_3\neq\infty, \\
		(x-x_1)^{c_{11}}(x-x_2)^{c_{21}}u(x), & \text{if } x_3=\infty,
	\end{cases}	
\end{equation}
brings the $P$-symbol in the form
\begin{equation}\label{key}
	P\begin{Bmatrix}x_1&x_2&x_3&\\
		0&0&c_{31}+c_{11}+c_{21}&x\\
		c_{12}-c_{11}&c_{22}-c_{21}&c_{32}+c_{11}+c_{21}&\end{Bmatrix}.
\end{equation}
The transformation,
\begin{equation}
	x \rightarrow \frac{(x_2-x_3)(x-x_1)}{(x_2-x_1)(x-x_3)},
\end{equation}
moves the locations of the singular points from ($x_1,x_2,x_3$) to ($0,1,\infty$) as in the standard form of the hypergeometric equation \cite{kris}. Now we have,
\begin{equation}\label{key}
	P\begin{Bmatrix}0&1&\infty&\\
		0&0&c_{31}+c_{11}+c_{21}&\frac{(x_2-x_3)(x-x_1)}{(x_2-x_1)(x-x_3)}\\
		c_{12}-c_{11}&c_{22}-c_{21}&c_{32}+c_{11}+c_{21}&\end{Bmatrix},
\end{equation}
which corresponds to the standard form of the hypergeometric equation. We note that the sum of the indices is not changed. A similar analysis of the general Heun equation can be found in \cite{kris}. 

Our approach is similar to this example for other equations: we change the indices of the regular singular points and move the locations of the singular points in order to obtain a standard form. After reaching the standard singularity structure of an equation that can be recognized by the code, the parameters are read either from the characteristic exponents or by matching the final form of the input equation with the standard equation in their normal forms. 

The polynomial coefficients of the normal forms are matched in the confluent cases. The parameters of these ODEs can be found by solving single equations, i.e. the code finds some parameters by solving algebraic equations that depend only on one parameter. The rest of the parameters are found by substitution. For the Fuschian ODEs, the parameters are read from the characteristic exponents, e.g. the non-zero exponent of the singular point at zero in the hypergeometric equation yields $1-c$, etc.

We find the parameters with this method and use the Maple or Mathematica forms of the solutions to substitute these parameters. The standard forms of the equations are given in the Appendix.

The hypergeometric equation has three pairs of Frobenius solutions around its three regular singular points and these solutions can be transformed into other solutions via specific transformations \cite{nist}. The number of all solutions of the hypergeometric equation is 24. The number of total solutions is 192 for the general Heun equation \cite{maier192}. The user of our code may need to use some transformations or function identities in order to obtain the desired form of the solution \cite{maier192,maierheuntohyper,nist}.

The results of the hypergeometric and confluent hypergeometric equations are numerically usable as these functions are defined in SageMath. However, the numerical solutions of the Heun-type functions are not defined. The numerical solutions of the general Heun and (singly) confluent Heun functions are defined by Motygin for GNU Octave/MATLAB \cite{oleg1,oleg2}. GNU Octave/MATLAB commands can be run in a SageMath session. However, this procedure is not straightforward and it is beyond the scope of this work. The method given in \cite{giscard} can also be employed in order to obtain numerical results. An optimized implementation of Giscard and Tamar's method is in progress, and the initial results of this work are presented in \cite{numericalheun}.

Several applications with symODE2 can be found in the User Manual and the sample worksheet of the code \cite{mygithub} based on the results obtained in \cite{Nasheeha:2020iym,Petroff:2007tz,Sakalli:2018nug,Vitoria:2015vee,Vieira:2015uua,Dong:2019wxa,ornek1,ornek2,ornek3,ornek4,dere,nist}.

\section{Conclusion}
We studied the singularity structures of the second-order ordinary differential equations with polynomial coefficients using symbolic analysis. Employing this information and appropriate calculations, we attempted to obtain solutions in terms of hypergeometric or Heun-type functions.

Using the theory, we proposed an open-source package under SageMath. Our approach was based on the singularity structure, namely, the locations of the singularities, corresponding characteristic exponents, and the ranks of the irregular singular points of the equation. Using particular substitutions and transformations, the singularity structure of the input equation was brought in a standard form that could be recognized by the routines. After being reached the standard singularity structure of an equation, the parameters were obtained either using the characteristic exponents or by matching the final form of the input equation with the normal form of the standard equation. 

The singularity structure, indices, and recurrence relations associated with the regular singular points, and symbolic solutions of the hypergeometric equation, Heun equation, and their confluent forms could be found using the package. 

As they were defined in SageMath, the results of the hypergeometric and confluent hypergeometric equations were numerically usable, unlike the Heun-type functions. 

We presented that our code worked properly with a number of tests.  We also mentioned that some transformations, substitutions, or identities might be needed in order to reach the results of the literature.

\section*{Acknowledgement} 
The author would like to thank Profs. Mahmut Horta\c{c}su, Nazmi Postacıo\u{g}lu, and Pierre-Louis Giscard for stimulating discussions. The author is also indebted to Maplesoft and Wolfram Research Inc. for providing excellent online resources for the Heun community.

\appendix
\section{Hypergeometric and Heun type equations in DLMF, Maple, Mathematica, and symODE2}
The standard forms of the equations may be defined differently in the literature and in the computer algebra systems. Here, we consider DLMF, a well-known library of mathematical functions \cite{nist}, and two computer algebra systems, Maple and Mathematica which can work with hypergeometric and Heun type functions. We also note the standard forms of the equations used in the symODE2 code.

In the Appendix of the User Manual, we also give a lists of correspondence of the parameters in these programs \cite{mygithub}. 
\subsection{Hypergeometric equation}
In DLMF \cite{nist},
\begin{equation}\label{dlmfhyper}
	x(1-x)\frac{d^2 y}{dx^2}+\left[c-(a+b+1)y\right]\frac{dy}{dx}-aby=0.
\end{equation}
In Maple \cite{maple2f1},
\begin{equation}
	x(x-1)\frac{d^2 y}{dx^2}+[(a+b+1)x-c]\frac{dy}{dx}+aby=0,
\end{equation}
with solution \verb|hypergeom([a,b],[c],x)|. 

In Mathematica \cite{mathematica2f1},
\begin{equation}
	-\bigg(x(x-1)\frac{d^2 y}{dx^2}+[(a+b+1)x-c]\frac{dy}{dx}+aby \bigg) =0,
\end{equation}
with solution \verb|Hypergeometric2F1[a,b,c,z]|. 

All equations coincide and they have three regular singularities located at \{0, 1, $\infty$\}. symODE2 uses this form of the equation as well.

\subsection{Confluent hypergeometric equation}
In DLMF \cite{nist},
\begin{equation}
	x\frac{d^2 y}{dx^2}+(b-x)\frac{dy}{dx}-ay=0.
\end{equation}
In Maple \cite{maple2f1},
\begin{equation}
	x\frac{d^2 y}{dx^2}+(c-x)\frac{dy}{dx}-ay=0,
\end{equation}
with solution \verb|hypergeom([a],[c],x)|.

In Mathematica \cite{mathematica1f1},
\begin{equation}
	x\frac{d^2 y}{dx^2}+(b-x)\frac{dy}{dx}-ay=0,
\end{equation}
with solution \verb|HypergeometricU[a,b,x]|.

All have one regular singularity located at \{0\} and one irregular singularity of rank-1 at \{$\infty$\}. The equations coincide and symODE2 also uses this form of the equation. 

\subsection{(General) Heun equation}
In DLMF \cite{nist},
\begin{equation}
	\frac{d^2 y}{dx^2}+\left(\frac{\gamma}{x}+\frac{%
		\delta}{x-1}+\frac{\epsilon}{x-a}\right)\frac{dy}{dx}+\frac{%
		\alpha\beta x-q}{x(x-1)(x-a)}y=0,
\end{equation}
and in Maple \cite{mapleheun},
\begin{equation}
	\frac{d^2 y}{dx^2}+\bigg(\frac{\gamma}{x}+\frac{\delta}{x-1}+\frac{\epsilon}{x-a} \bigg)\frac{dy}{dx}+\frac{\alpha \beta x -q}{x(x-1)(x-a)}y=0,
\end{equation}
where $\epsilon=\alpha+\beta+1-\gamma-\delta$ with solution \verb|HeunG|$(a,q,\alpha,\beta,\gamma,\delta,x)$.

In Mathematica \cite{mathematicaheun},
\begin{equation}
	\frac{d^2 y}{dx^2}+\bigg(\frac{\gamma}{x}+\frac{\delta}{x-1}+\frac{\alpha+\beta+1-\gamma-\delta}{x-a} \bigg)\frac{dy}{dx}+\frac{\alpha \beta x -q}{x(x-1)(x-a)}y=0,
\end{equation}
with solution \verb|HeunG|$(a,q,\alpha,\beta,\gamma,\delta,x)$.

All have four regular singularities located at \{0, 1, $a$, $\infty$\}. symODE2 uses the same structure of the equation.

\subsection{(Singly) Confluent Heun equation}
In DLMF \cite{nist},
\begin{equation}
	\frac{d^2 y}{dx^2}+\left(\frac{\gamma}{x}+\frac{\delta}{x-1}+\epsilon\right)\frac{dy}{dx}+\frac{\alpha x-q}{x(x-1)}y=0.
\end{equation}
In Maple \cite{mapleheun},
\begin{eqnarray}
	&&\frac{d^2 y}{dx^2}
	-{\frac {-{x}^{2}\alpha+ \left( -\beta+\alpha-\gamma-2 \right) x+\beta+1}{x \left( x-1 \right) }}
	\frac{dy}{dx} \nonumber\\
	&&-{\frac { \left[  \left( -\beta-\gamma-2 \right) \alpha-2\,\delta
			\right] x+ \left( \beta+1 \right) \alpha+ \left( -\gamma-1 \right) 
			\beta-2\,\eta-\gamma}{2x \left( x-1 \right) }}y=0,
\end{eqnarray}
with solution \verb|HeunC|$( \alpha,\beta,\gamma,\delta,\eta,x)$. This form can be transformed into
\begin{equation}
	\frac{d^2 y}{dx^2}+\bigg(\frac{\beta+1}{x}+\frac{\gamma+1}{x-1}+\alpha \bigg)\frac{dy}{dx}+\bigg(\frac{\mu}{x} + \frac{\nu}{x-1} \bigg)y=0,	
\end{equation}
where
\begin{eqnarray}
	\delta&=&\mu+\nu-\alpha \frac{\beta+\gamma+2}{2},\\
	\eta&=&\frac{(\alpha-\gamma)(\beta+1)-\beta}{2}-\mu.
\end{eqnarray}
In Mathematica \cite{mathematicaheun},
\begin{equation}\label{mathheunc}
	\frac{d^2 y}{dx^2}+\bigg(\frac{\gamma}{x}+\frac{\delta}{x-1}+\epsilon \bigg)\frac{dy}{dx}+\frac{\alpha x -q}{x(x-1)}y=0,
\end{equation}
with solution \verb|HeunC|$(q,\alpha,\gamma,\delta,\epsilon,x)$. 


All have two regular singularities located at \{0,1\} and an irregular singularity of rank-1 at \{$\infty$\}. symODE2 uses the Maple form.

\subsection{Double confluent Heun equation}
In DLMF \cite{nist},
\begin{equation}
	\frac{d^2 y}{dx^2}+\left(\frac{\delta}{x^{2}}+\frac{\gamma}{x}+1\right)\frac{dy}{dx}+\frac{\alpha x-q}{x^{2}}y=0.
\end{equation}
This equation has two irregular singular points of rank-1 located at \{0, $\infty$\}.

In Maple \cite{mapleheun},
\begin{equation}
	\frac{d^2 y}{dx^2}
	-{\frac {\alpha\,{x}^{4}-2\,{x}^{5}+4\,{x}^{3}-\alpha-2\,x}{ \left( x-
			1 \right) ^{3} \left( x+1 \right) ^{3}}}
	\frac{dy}{dx}
	-{\frac {-{x}^{2}\beta+ \left( -\gamma-2\,\alpha \right) x-\delta}{
			\left( x-1 \right) ^{3} \left( x+1 \right) ^{3}}}
	y=0,
\end{equation}
with solution \verb|HeunD|$( \alpha,\beta,\gamma,\delta,x)$. This equation has two irregular singular points of rank-1 located at \{-1, 1\}.

In Mathematica \cite{mathematicaheun},
\begin{equation}\label{mathheund}
	\frac{d^2 y}{dx^2}+\bigg(\frac{\gamma}{x^2}+\frac{\delta}{x}+\epsilon \bigg)\frac{dy}{dx}+\frac{\alpha x -q}{x^2}y=0,
\end{equation}
with solution \verb|HeunD|$(q,\alpha,\gamma,\delta,\epsilon,x)$. This equation has two irregular singular points of rank-1 located at \{0, $\infty$\}. symODE2 uses the Mathematica form.

\subsection{Biconfluent Heun equation}
In DLMF \cite{nist},
\begin{equation}
	\frac{d^2 y}{dx^2}-\left(\frac{\gamma}{x}+\delta+x\right)\frac{dy}{dx}+\frac{\alpha x-q}{z}y=0.
\end{equation}
In Maple \cite{mapleheun},
\begin{equation}
	\frac{d^2 y}{dx^2}
	-{\frac {\beta\,x+2\,{x}^{2}-\alpha-1}{x}}
	\frac{dy}{dx}
	-{\frac { \left( 2\,\alpha-2\,\gamma+4 \right) x+\beta\,\alpha+
			\beta+\delta}{2x}}
	y=0,
\end{equation}
with solution \verb|HeunB|$( \alpha,\beta,\gamma,\delta,x)$. 

In Mathematica \cite{mathematicaheun},
\begin{equation}\label{mathheunb}
	\frac{d^2 y}{dx^2}+\bigg(\frac{\gamma}{x}+\delta+\epsilon x  \bigg)\frac{dy}{dx}+\frac{\alpha x -q}{x}y=0,
\end{equation}
with solution \verb|HeunB|$(q,\alpha,\gamma,\delta,\epsilon,x)$. 

All have one regular singularity located at \{0\} and one irregular singularity of rank-2 at \{$\infty$\}. symODE2 uses the Mathematica form.


\subsection{Triconfluent Heun equation}
In DLMF \cite{nist},
\begin{equation}
	\frac{d^2 y}{dx^2}+\left(\gamma+x\right)x\frac{dy}{dx}+\left(\alpha x-q\right)y=0.
\end{equation}
In Maple \cite{mapleheun},
\begin{equation}
	\frac{d^2 y}{dx^2}
	-(3{x}^{2}+\gamma)
	\frac{dy}{dx}
	+ [\left( \beta-3 \right) x+\alpha]
	y=0,
\end{equation}
with solution \verb|HeunT|$( \alpha,\beta,\gamma,x)$. 

In Mathematica \cite{mathematicaheun},
\begin{equation}
	\frac{d^2 y}{dx^2}+(\gamma + \delta x + \epsilon x^2)
	\frac{dy}{dx}+(\alpha x-q)y=0,
\end{equation}
with solution \verb|HeunT|$(q,\alpha,\gamma,\delta,\epsilon,x)$.

All have one irregular singularity of rank-3 at \{$\infty$\}. symODE2 uses the Mathematica form.


\label{LastPage}

\end{document}